\documentclass[12pt]{iopart}
\usepackage{epsfig}
\usepackage{iopams}
\begin{document}

\title[Probing the quantum coherence of a nanomechanical resonator: Echo scheme]
{Probing the quantum coherence of a nanomechanical resonator using a superconducting qubit: I. Echo scheme}% Force line breaks with \\
\author{A.D. Armour$^{\dag}$ and M.P. Blencowe$^{\ddag}$}
\address{\dag School of Physics and Astronomy, University of Nottingham,
Nottingham,\\ NG7 2RD, UK\\
\ddag Department of Physics and Astronomy, 6127 Wilder Laboratory,
Dartmouth College, Hanover, NH 03755, USA}
\ead{\mailto{andrew.armour@nottingham.ac.uk}\\\mailto{miles.p.blencowe@dartmouth.edu}}

\begin{abstract}
We propose a scheme in which the quantum coherence of a
nanomechanical resonator can be probed using a superconducting
qubit. We consider a mechanical resonator coupled capacitively to a
Cooper-pair box and assume that the superconducting qubit is tuned
to the degeneracy point so that its coherence time is maximised and
the electro-mechanical coupling can be approximated by a dispersive
Hamiltonian. When the qubit is prepared in a superposition of states
this drives the mechanical resonator progressively into a
superposition which in turn leads to apparent decoherence of the
qubit. Applying a suitable control pulse to the qubit allows its
population to be inverted resulting in a reversal of the resonator
dynamics. However, the resonator's interactions with its environment
mean that the dynamics is not completely reversible. We show that
this irreversibility is largely due to the decoherence of the
mechanical resonator and can be inferred from appropriate
measurements on the qubit alone. Using estimates for the parameters
involved based on a specific realization of the system we show that
it should be possible to carry out this scheme with existing device
technology.
\end{abstract}

\submitto{NJP} \maketitle

\section{Introduction}

One way of exploring the quantum coherence properties of a
nanomechanical resonator is to couple it to a qubit formed by a
solid-state two-level system (TLS). Coupling to an isolated harmonic
oscillator can initially cause an {\it apparent} loss of phase
coherence in the qubit if the oscillator is driven into a
superposition of orthogonal states, but signatures of the overall
coherence of the full system (i.e.\ oscillator and TLS together) can
be found in the subsequent dynamics of the TLS. However, if instead
the qubit is coupled to a harmonic oscillator which is in turn
coupled to a bath, then the effective dynamics of the TLS and
oscillator will now be different and the loss of the oscillator's
coherence due to the bath will be manifest in the dynamics of the
TLS\,\cite{ABS,Buks,Tian}.

From a theoretical point of view it is relatively straightforward to
devise simple schemes based on these principles to probe the rate at
which the environment causes decoherence of an
oscillator\,\cite{deconstruct}. Indeed, exactly this kind of
approach has been used very successfully in the field of cavity
quantum electrodynamics (cQED) to probe the quantum coherence of a
mode of the electromagnetic field by examining its influence on
effectively two-level atoms\,\cite{ramsey,spinex,Haroche}. Similar
experiments have also been carried out successfully on trapped ions
with the internal electronic state of the ion playing the role of
the two-level system and the ion's motional state the
oscillator\,\cite{monroe,leibfried,mcdonnell}.

The development of relatively large and well controlled quantum
coherent TLSs in the solid-state, such as superconducting circuits
designed to act as qubits, seems to offer a way to perform analogous
experiments with nanomechanical resonators\,\cite{ABS,Buks}.
Furthermore, recent experiments have demonstrated that it is
possible to recreate many of the features of traditional optical
cQED in the solid state using a superconducting qubit coupled to a
superconducting resonator\,\cite{Yale1,phase}. Since nanomechanical
resonators are typically a few microns in length and contain
macroscopic numbers of atoms, producing a quantum superposition of
spatially separated states in such systems and monitoring its
progressive loss of coherence (due to interactions with its
environment) would represent an important increase in size of the
system involved compared to superpositions of ions and
light\,\cite{mpb,leggett}. However, performing quantum coherent
experiments using a nanomechanical resonator is likely to be more
difficult than with a superconducting one as nanomechanical
resonators are generally much lower in frequency.

In order to understand the practical difficulties entailed in using
a superconducting qubit to probe the decoherence of a nanomechanical
resonator we briefly review the apparent constraints which any such
scheme must satisfy. First of all, the superconducting qubit must
remain sufficiently coherent that the influence of the {\it
mechanical resonator's environment} can be clearly discerned in its
dynamics. Secondly, it will be desirable to couple the TLS and
resonator as strongly as possible since the signal(s) of coherence
and/or decoherence in the mechanical resonator measurable in the TLS
will become clearer the larger the coupling between the TLS and
resonator. Finally, unless impressive cooling of the resonator can
be achieved, the experiments will always have to contend with the
competing effect of phase smearing arising from the range of
oscillator states (and their associated phases) in the thermal
ensemble of the oscillator. Again the relatively low frequencies of
mechanical resonators make this more of a problem than it would be
in the superconducting case. Note that in practice there is no
simple way of designing a double clamped beam resonator to optimize
all of these constraints at once\footnote{One interesting way to
avoid the problems posed by the relatively low frequency of
flexural-mode mechanical resonators is to use a different type of
mechanical system, such as dilational disk resonators which can have
frequencies well beyond 1 GHz\,\cite{Cleland}. Here, however, we will
confine our attention to the conventional mechanical resonators
formed by doubly clamped beams.}. For example, in most realizations
the TLS-resonator coupling will increase as the resonator is made
larger, but enlargement of the resonator will inevitably reduce its
fundamental frequency.

In this paper we describe how a dispersive interaction between a
superconducting qubit and a nanomechanical resonator can be used to
produce superpositions of the resonator state and how the coherence
of this superposition can then be probed by measuring the state of
the TLS. We identify the relatively short coherence times of the
superconducting qubit as the most serious constraint on these types
of scheme and hence assume that the TLS is tuned to operate at a
point where its coherence is maximized; it is this choice of
operating point which leads to a dispersive coupling between the TLS
and the resonator. Although the dispersive interaction is relatively
weak, we find that the effect of the TLS on the resonator can be
amplified by preparing the latter in a state with large amplitude.
We explore in detail the quantum dynamics of the resonator and TLS
including the effects of the inevitably mixed initial state of the
resonator and the interaction with the environment. In assessing the
extent to which the schemes we propose can be carried out in
practice we make use of the analysis in the companion reference\ \cite{Blencowe}, which
considers how sufficiently strong coupling between a nanomechanical resonator and
a superconducting qubit can be best   achieved without degrading
the coherence of the qubit.

Our work builds on and extends previous studies of similar
systems\,\cite{ABS,Buks,pert,Tian,ringsmuth}. In particular, we
believe that the scheme outlined here represents an important
improvement on that proposed by us in Ref.\ \cite{ABS} in a number
of respects. Most importantly, the scheme we propose here is
more likely to be practicable as it is designed to be performed at
the degeneracy point of the qubit where it remains coherent for at
least an order of magnitude longer\,\cite{qubitcoherence} than the
operating point considered in \cite{ABS}. Furthermore, the scheme is
much more flexible in the sense that it would be possible to vary
several of the important parameters systematically (such as the
phase space separation of the resonator states involved and the
duration of the superposition). This would be an important advantage
in interpreting the results of this type of experiment, since the
nature of the mechanical resonator's environment is not well
understood \cite{schloss} and in a sense the purpose of the
experiment we propose would be to provide empirical information
about it.

This paper is organised as follows. In section \ref{Sec:2} we
introduce the generic Hamiltonian for the superconducting
qubit-resonator systems which we will work with here. We discuss the
practical constraints which dictate our choice of operating regime
and introduce the effective (dispersive) Hamiltonian which is valid
when the mechanical resonator is much slower than the
superconducting circuit. Next in section \ref{Sec:simple} we
describe how the dispersive interaction can create states which
involve superpositions of spatially separated states of the
mechanical resonator so that measurements on the TLS alone show an
apparent initial loss of coherence. We show that the TLS coherence
can be recovered (recoherence) in a controlled way using a
particular choice of control pulses. In section \ref{Sec:env} we
calculate how the presence of the environment of both the mechanical
resonator and the qubit itself affects recoherence. We then consider
the values of the various parameters which are likely to be
practicable in present or near future experiments in section\
\ref{sec:prac}. Then in section \ref{Sec:res}, we present
calculations of the behaviour of the recoherences for a range of
practicable parameter values. Finally, section \ref{Sec:conc}
contains a discussion of our results and our conclusions. The
Appendix contains further details on some of our calculations.

\section{Resonator-TLS Effective Hamiltonian}
\label{Sec:2}

\subsection{Operating Regime}
The generic Hamiltonian for the superconducting TLS and mechanical
resonator which we consider here is,
\begin{equation}
H=\frac{\epsilon_0}{2}\sigma_z+\Delta
\sigma_x+\lambda(a^{\dagger}+a)\sigma_z+\hbar\omega
\left(a^{\dagger}a +\frac{1}{2}\right)\label{H0}
\end{equation}
where the qubit energy scales $\epsilon_0$ and $\Delta$ depend on
the details of the specific superconducting system considered,
$\omega$ is the resonator frequency, $\lambda$ is the strength of
the resonator-TLS coupling, and the operators $\sigma_{z(x)}$ act on
the TLS while $a^{(\dagger)}$ act on the resonator. The mechanical
resonator is assumed to be the fundamental flexural mode of a
suspended doubly-clamped beam. The coupling between the TLS and the
mechanical resonator is implicitly assumed to be weak in the sense
that only linear (in the resonator position coordinate) coupling
needs to be considered. The TLS states are defined as $|1\rangle$
and $|0\rangle$ so that e.g. $\sigma_x=|1\rangle\langle
0|+|0\rangle\langle 1|$.

This Hamiltonian is derived in Ref.\ \cite{Blencowe} for the
specific cases of either a Cooper pair box (CPB) with an island
which is suspended to form the mechanical resonator, or a flux qubit
(again with suspended segment that forms the mechanical resonator).
In each case the qubit is also assumed to be fabricated close to the
centre electrode of a superconducting microwave coplanar waveguide (CPW) resonator. The
CPW resonator provides a way to both measure and
manipulate the qubit state, as well as a means to drive the mechanical resonator into an initial state
which has a large amplitude. Although both the mechanical resonator
and the qubit are also coupled to the CPW
resonator in this system as just stated, we will not include the latter explicitly in this article, assuming that it is
unpopulated (i.e. it is at or close to the vacuum state), except
initially when used to drive the mechanical resonator\,\cite{Blencowe}
and for the short periods when it is used to manipulate the state of
the qubit.
Population of the CPW resonator is required when measuring the qubit
state, but at this stage disruption to the mechanical system and
dephasing of the qubit are unimportant so long as the measurement
can still be performed with high fidelity, which we will assume is
the case. For a CPB the states $|0\rangle,|1\rangle$ correspond to
different charge states and the coupling between the TLS and the
resonator is capacitive. In contrast, when the Hamiltonian [equation
(\ref{H0})] is realized with a flux qubit, the coupling between the
TLS and the mechanical resonator is inductive and the relevant qubit
states are of current circulating in opposite directions\ \cite{Buks,Blencowe,flux1}.

 The best coherence times for both superconducting
charge and flux based qubits  are achieved at the
degeneracy point where $\epsilon_0=0$. Away from the degeneracy
point, experiments\,\cite{qubitcoherence} have demonstrated that the
coherence times of superconducting qubits decrease by orders of
magnitude. We regard the coherence of the superconducting TLS as the
primary constraint and hence we choose to operate the TLS at its
degeneracy point when probing the resonator's coherence. Another
important constraint arising from the use of the superconducting TLS
is the need to avoid thermal mixing of the two states involved. In
practice, for experiments performed at temperatures of order 20~mK
this means that we will require energy separations between the two
states (i.e. $2\Delta$ when working at the degeneracy point) in the
superconducting TLS that are much larger than the thermal energy scale.
Experiments\,\cite{Yale2005} using a CPB (coupled to a
superconducting CPW resonator for state control and read-out)
achieved coherence ($T_2$) times of up to $0.5~\mu$s (operating at
the degeneracy point) and relaxation times ($T_1$) of about $7~\mu$s
with $\nu_a=2\Delta/h\sim 5$~GHz and we take these as indicative of
the current practical limitations. Note that even longer coherence
times of up to $2~\mu$s have been reported for experiments on CPBs
using an echo technique in which the TLS state is inverted mid-way
through the experiment\cite{Wallraff2007}

In terms of the mechanical resonator, we will consider beam
structures which are fabricated by under-etching a bulk substrate or
 metallic film. The fundamental (flexural) mode frequencies of such devices
can in practice be as high as $1$~GHz\,\cite{Huang}, but because the
electro-mechanical coupling, $\lambda$, (in equation \ref{H0}) for
such modes increases with the length of the beam, $l$, (for both
capacitive and inductive couplings\,\cite{Blencowe}) whereas the
frequency clearly decreases with increasing $l$, it is clear that
high frequencies can only be achieved at the expense of very weak
couplings\,\cite{Blencowe,ABS,Buks}. Nanomechanical resonators have
already been fabricated in close proximity to superconducting
structures\,\cite{Naik}, but with mechanical frequencies $\sim
20$~MHz. We therefore assume that the mechanical frequency will be
much lower than the energy scale of the qubit, i.e.\ $\Delta\gg
\hbar\omega$. Having made this assumption of a wide separation of
time-scales for the mechanical and superconducting elements we can
proceed to derive a simpler effective Hamiltonian which is valid in
this regime.\footnote{Note that the CPW resonator
used to manipulate the CPB state will have a frequency which is
close to $\nu_a$ and hence is also very far from the mechanical frequency.}

\subsection{Adiabatic limit}

When  $\epsilon_0$, is tuned to zero it is convenient to rewrite the
Hamiltonian in terms of a new basis for the qubit. Defining new
basis states,
\begin{equation}
|\pm\rangle=\frac{1}{\sqrt{2}}(\pm|0\rangle+|1\rangle),
\end{equation}
we can write the Hamiltonian (\ref{H0}) for $\epsilon=0$ as
\begin{equation}
H_{\rm deg}=\Delta
\overline{\sigma}_z+\lambda(a^{\dagger}+a)\overline{\sigma}_x+\hbar\omega
\left(a^{\dagger}a +\frac{1}{2}\right)\label{H1}
\end{equation}
where the new spin operators are defined in terms of the new basis
states.

We proceed by exploiting the separation in time-scales to make an
adiabatic approximation\,\cite{Buks,larson,messiah,graham}. Since
the mechanical resonator will generally be in a Gaussian state of
large amplitude, in what follows it is reasonable to take a
semi-classical approach\,\cite{larson}. We initially assume that the
mechanical resonator is at a fixed position $x$ and use this to
calculate the eigenvalues of the TLS, these are then used to
calculate an effective Hamiltonian for the oscillator. With the
resonator at position $x$ the Hamiltonian of the TLS is (using
equation \ref{H1}):
\begin{equation}
H_{TLS}=\Delta\overline{\sigma}_z+\lambda
(x/x_{zp})\overline{\sigma}_x, \label{tls}
\end{equation}
where $x_{zp}=(\hbar/2m\omega)^{1/2}$. The eigenvalues of the TLS
are now $\epsilon_{\pm}=\pm \sqrt{\Delta^2+(\lambda x /x_{zp})^2}$.
For sufficiently weak coupling i.e. $[\lambda x
/(x_{zp}\Delta)]^2\ll 1$, we can expand the eigenvalues to lowest
order
\begin{equation}
\epsilon_{\pm}=\pm\Delta\left(1+\frac{1}{2}\left(\frac{\lambda
x}{x_{zp}\Delta}\right)^2\right). \label{eps}
\end{equation}
The evolution of the mechanical system over time then causes a weak
position dependent (and hence ultimately time dependent)
perturbation to the eigenstates of the TLS. In the adiabatic
approximation the wide separation of time-scales and weak coupling
mean that the TLS evolves smoothly {\it within} each eigenstate with
its dynamics arising from changes in time of the eigenstates
themselves, rather than any transitions between different
eigenstates.

Labeling the instantaneous eigenstates of equation (\ref{tls}) as
$|\tilde{+}\rangle$ and $|\tilde{-}\rangle$, we can write down the
effective Hamiltonian felt by the oscillator for the TLS confined to
one of its eigenstates as,
\begin{equation}
H_{\tilde{\pm}}=\hbar\omega\left(a^{\dagger}a +\frac{1}{2}\right)\pm
\left(\Delta+\frac{\lambda^2}{2\Delta}(a^\dagger+a)^2\right)
\end{equation}
Therefore within the framework of the adiabatic approximation we can
write down the following model Hamiltonian for the system,
\begin{equation}
H=\Delta\left(1
+\frac{\lambda^2}{2\Delta^2}(a^{\dagger}+a)^2\right)\overline{\sigma}_z+\hbar\omega
\left(a^{\dagger}a +\frac{1}{2}\right)\label{H_dis}.
\end{equation}
We have dropped the distinction between perturbed and unperturbed
eigenstates as it does not play a role in what follows.

Assuming that the coupling term is a weak perturbation  (i.e.\
assuming $\lambda^2/2\Delta\ll \hbar\omega$), we can also make the
rotating wave approximation in which the terms $a^2$ and
$(a^{\dagger})^2$ are dropped\,\cite{Buks}. The final result of
these approximations is the following dispersive Hamiltonian for the
TLS-resonator system,
\begin{equation}
H_d=\Delta
\overline{\sigma}_z+\hbar\omega_1\overline{\sigma}_z\left(a^{\dagger}a+\frac{1}{2}\right)+\hbar\omega
\left(a^{\dagger}a+\frac{1}{2}\right) \label{dispersive}
\end{equation}
where $\omega_1=\lambda^2/(\hbar\Delta)$. A key feature of this
Hamiltonian is that the perturbation of the oscillator commutes with
the unperturbed Hamiltonian (i.e.\ it is a QND Hamiltonian). This
feature is exploited in schemes to measure the number state of a
resonator using a superconducting
qubit\,\cite{gambetta,Clerk,Buksqnd} and it also plays an important
role in what follows here. Note that this Hamiltonian can also be
obtained from equation (\ref{H1}) via a range of other
approaches\,\cite{Haroche,pert,Blais,Clerk,gerry}.

\section{Coherent oscillations and recoherences: simple description}
\label{Sec:simple}

\begin{figure}[t]
\centering \epsfig{file=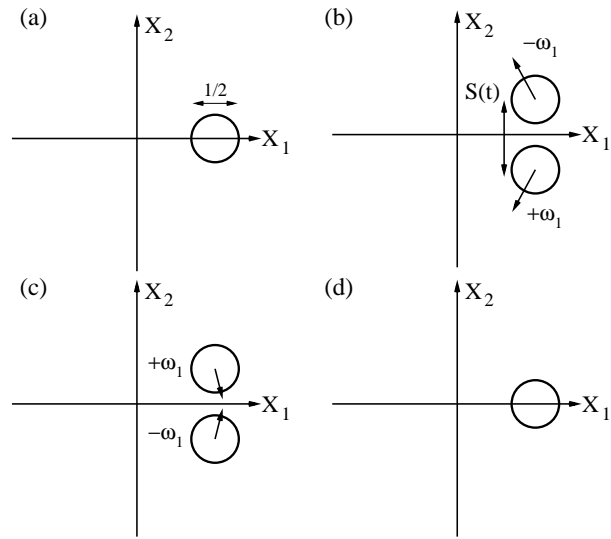,width=8cm} \caption{Schematic
illustration of the evolution of the mechanical resonator in phase
space during the echo sequence. Initially (a) the resonator is
prepared in a coherent state and the qubit is prepared in a
superposition of states. The two qubit states couple to the
resonator leading to different effective frequencies $\omega\pm
\omega_1$ so that in the frame rotating at the resonator frequency
the two mechanical states start to pull apart (b). A $\pi$ pulse
inverts the qubit state and hence interchanges the relative
frequencies of the two resonator states (c). When the periods of
evolution before and after the inversion of the qubit are the same
the resonator will return to its initial state (d) {\it in the
absence of dissipation}. } \label{fig:f0}
\end{figure}

The dispersive Hamiltonian shifts the mechanical frequency in a way
that depends on the state of the TLS. This interaction can be used
to probe the quantum coherence of the mechanical resonator. The idea
is to perform a Ramsey interference\,\cite{Haroche} experiment in
which the qubit is prepared in a superposition of its eigenstates
states using a control pulse, this superposition is then allowed to
interact with the resonator for a time $t$ before a second pulse is
applied to the qubit and then a measurement of its state is
performed. For an isolated TLS the probability of finding the system
in one or other of its eigenstates at the end of the experiment will
oscillate between zero and unity as a function of the time between
the two control pulses. When the mechanical resonator is present the
interaction with the superposition of TLS states leads to an overall
superposition of states involving spatially separated mechanical
states. For a sufficiently strong interaction, the separation of the
resonator states coupled to the qubit states leads to a strong
suppression of the oscillations in the final qubit state
measurements. The coherence of the resonator can be inferred by
inverting the state of the TLS midway between the two original
control pulses. The scheme is illustrated schematically in figure
\ref{fig:f0}. In the absence of the resonator's environment, such an
inversion should lead to a reversal of its dynamics and hence the
recovery of the oscillations in the final TLS state
measurement\,\cite{spinex}. Very similar schemes have been
demonstrated in optical systems\,\cite{Haroche}.

In what follows, we will assume that it is possible to measure the
state of the TLS within the $|\pm\rangle$ basis and to rotate its
state by applying transformations of the form
$\exp(-i\theta\overline{\sigma}_x/2)$ with a parameter $\theta$ that
can be controlled to a high degree of precision. These requirements
are readily met in the system of a charge or flux qubit (with
suspended segment forming a mechanical resonator) coupled to a
superconducting CPW resonator described in Ref.\ \cite{Blencowe},
which we consider here. The TLS state is determined by measuring the
transmission of an off-resonant pulse applied to the CPW resonator,
while rotation of the TLS is performed by applying almost-resonant
pulses to the CPW resonator and making use of the resulting Rabi
oscillations\,\cite{Blais,Yale2005}.

 We will begin by
considering the simple though unrealistic case of an isolated
resonator which is initially prepared in a coherent state,
$|\alpha_0\rangle$. The effects of the environment on the evolution
of the coupled TLS-resonator device and the types of initial
(mixtures of) states of the oscillator that can be prepared in
practice are addressed in later sections. We assume that the TLS
system is in its ground state $|-\rangle$, hence the total initial
state is $|-\rangle\otimes|\alpha_0\rangle$. Application of an
appropriate control pulse to rotate the state of the TLS by
$\theta=\pi/2$ produces a superposition of TLS states. Since the
rotation of the TLS will in practice be very fast compared to the
mechanical period we can neglect any evolution of the mechanical
resonator during the pulse and hence write the total state of the
system after the pulse as $\rho (0)=|\psi(0)\rangle\langle \psi(0)|$
with
\begin{equation}
|\psi(0)\rangle=\frac{1}{\sqrt{2}}(|-\rangle-i|+\rangle)\otimes|\alpha_0\rangle,
\label{psi0}
\end{equation}
Starting with this initial state at $t=0$, the dispersive
interaction [equation (\ref{dispersive})] leads to the following joint
state after time $t$,
\begin{equation}
|\psi(t)\rangle=\frac{1}{\sqrt{2}}\left(|-\rangle\otimes|\alpha_-(t)\rangle-i|+\rangle\otimes|\alpha_+(t)\rangle\right)
\end{equation}
where
\begin{eqnarray}
|\alpha_{\pm}(t)\rangle&=&{\mathrm e}^{\mp i\Delta t/\hbar}{\mathrm
e}^{-i(\omega\pm\omega_1)(a^{\dagger}a+1/2)t}|\alpha_0\rangle\\
&=&{\mathrm e}^{\mp i\Delta
t/\hbar}\mathrm{e}^{-i(\omega\pm\omega_1)t/2}|\alpha_0\mathrm{e}^{-i(\omega\pm\omega_1)t}\rangle.
\label{alphapm}
\end{eqnarray}
The resonator evolution in phase space during this period is
illustrated in figure \ref{fig:f0}b. The next step is to perform a
second $\pi/2$ rotation on the TLS, leading to the state
\begin{equation}
|\psi^{(+)}(t)\rangle=\frac{1}{2}\left[|-\rangle\otimes\left(|\alpha_-(t)\rangle-|\alpha_+(t)\rangle\right)-i|
+\rangle\otimes\left(|\alpha_-(t)\rangle+|\alpha_+(t)\rangle\right)\right].
\end{equation}

Finally, the state of the TLS is measured in the $|\pm\rangle$
basis. The probability of finding the TLS in state $|+\rangle$ for a
period of evolution $t$  between the two control pulses is
\begin{equation}
P_{|+\rangle}(t)={\rm
Tr}[|+\rangle\langle+|\rho(t)]=\frac{1}{2}\left(1+{\rm Re}[\langle
\alpha_-(t)|\alpha_+(t)\rangle]\right).
\end{equation}
The overlap is readily evaluated,
\begin{equation}
\langle \alpha_-(t)|\alpha_+(t)\rangle ={\rm
e}^{-|\alpha_0|^2\left(1-{\rm e}^{-2i\omega_1t}\right)}{\rm
e}^{-2i\Delta t/\hbar-i\omega_1t}.
\end{equation}
The final result for $P_{|+\rangle}(t)$ is thus,
\begin{equation}
P_{|+\rangle}(t)=\frac{1}{2}+\frac{1}{2}{\rm Re}\left[{\rm
e}^{-|\alpha_0|^2\left(1-{\rm e}^{-2i\omega_1t}\right)}{\rm
e}^{-2i\Delta t/\hbar-i\omega_1t}\right]. \label{p0}
\end{equation}
Note that this function depends only on the amplitude of the initial
mechanical state, not its phase.

The behaviour of $P_{|+\rangle}(t)$ is easy to understand. Without
any coupling to the resonator the coherent oscillations in the TLS
state mean that the probability oscillates over time between zero
and unity with a period $\tau_{R}=h/(2\Delta)$\footnote{Note that in
practice the pulses used to rotate the state of the TLS are chosen
to be slightly off-resonant. As a result a stroboscopic observation
of the oscillations in $P_{|+\rangle}$ can be made which replaces
the very fast oscillations at frequency $2\Delta/\hbar$ with much
slower (and hence easier to observe) ones at the chosen de-tuning
frequency\,\cite{Haroche,vion,Yale2005}, we neglect this detail here
as our primary interest is not in the frequency of the oscillations
but in their amplitude.} a key indicator of the quantum coherence of
the TLS\,\cite{Yale2005}. For sufficiently strong coupling, the
resonator causes a relatively rapid reduction in the amplitude of
the oscillations as a function of time leading to a period where
$P_{|+\rangle}(t)=0.5$, implying that the resonator decoheres the
TLS. However, because the resonator is a periodic system and is
itself coherent, the oscillations in $P_{|+\rangle}(t)$ reappear,
giving rise to so-called recoherence, for $t\sim
\pi/\omega_1$\,\cite{Buks}.

Although recoherence does occur naturally after a time $t\sim
\pi/\omega_1$ it is preferable to use an approach where the time
between coherences can be varied systematically.  This is readily
achieved using a spin echo technique to induce recoherence at a
chosen time. This type of approach was used with great success in
optical cQED experiments\,\cite{spinex} as well as experiments on
superconducting circuits \,\cite{phase,Wallraff2007}.

\subsection{Echo technique}
For the spin echo sequence we again start with the system in the
state $|\psi(0)\rangle$ (equation \ref{psi0}) and allow it to evolve
as before for a time $t_1$ so that,
\begin{equation}
|\psi(t_1)\rangle=\frac{1}{\sqrt{2}}\left(|-\rangle\otimes|\alpha_-(t_1)\rangle-i|+\rangle\otimes|\alpha_+(t_1)\rangle\right).
\end{equation}
Next we apply a control pulse to the TLS which effectively inverts
the populations of the two eigenstates [this corresponds to the
unitary operation $\exp(-i\theta\overline{\sigma}_x/2)$ with
$\theta=\pi$]. Thus, just after the pulse we have
\begin{equation}
|\psi(t_1^+)\rangle=\frac{-1}{\sqrt{2}}\left(i|+\rangle\otimes|\alpha_-(t_1)\rangle+|-\rangle\otimes|\alpha_+(t_1)\rangle\right).
\end{equation}
We now allow the system to evolve for a further time $t_2$, after
which the resulting state of the system will be
\begin{equation}
|\psi(t_1+t_2)\rangle=\frac{-1}{\sqrt{2}}\left(|-\rangle\otimes|\alpha_{-+}(t_2,t_1)\rangle+i|+\rangle\otimes|\alpha_{+-}(t_2,t_1)\rangle\right).
\end{equation}
where now,
\begin{eqnarray}
|\alpha_{-+}(t_2,t_1)\rangle&=&{\mathrm e}^{-i\Delta
(t_1-t_2)/\hbar}{\mathrm
e}^{-i(\omega-\omega_1)(a^{\dagger}a+1/2)t_2}{\mathrm
e}^{-i(\omega+\omega_1)(a^{\dagger}a+1/2)t_1}|\alpha_0\rangle \\
|\alpha_{+-}(t_2,t_1)\rangle&=&{\mathrm e}^{i\Delta
(t_1-t_2)/\hbar}{\mathrm
e}^{-i(\omega+\omega_1)(a^{\dagger}a+1/2)t_2}{\mathrm
e}^{-i(\omega-\omega_1)(a^{\dagger}a+1/2)t_1}|\alpha_0\rangle.
\end{eqnarray}
Note that the simplicity of this expression relies on the fact that
the perturbed resonator Hamiltonian commutes with the unperturbed
one\footnote{Note, however, that with an appropriate choice of
control pulse at $t=t_1$ the idea of an echo experiment is not
limited to systems with dispersive Hamiltonians. This type of
experiment has been performed for systems with a Jaynes-Cummings
type interaction\,\cite{spinex}.}, thus we find
\begin{eqnarray}
\langle \alpha_{-+}(t_2,t_1)|\alpha_{+-}(t_2,t_1)\rangle &=&{\mathrm
e}^{2i\Delta (t_1-t_2)/\hbar}\langle \alpha_0|{\mathrm
e}^{i2\omega_1(a^{\dagger}a+1/2)(t_1-t_2)}|\alpha_0\rangle.
\end{eqnarray}
Carrying out a final rotation of the TLS state (with $\theta=\pi/2$)
the final overall probability of finding it in state $|+\rangle$ is
given by
\begin{equation}
P_{|+\rangle}(t_1+t_2)=\frac{1}{2}-\frac{1}{2}{\mathrm{Re}}\left[{\rm
e}^{-|\alpha_0|^2\left(1-{\rm e}^{2i\omega_1(t_1-t_2)}\right)}{\rm
e}^{2i\Delta (t_1-t_2)/\hbar+i\omega_1(t_1-t_2)}\right].
\end{equation}
The probability $P_{|+\rangle}$ is zero at $t=t_1+t_2$ when
$t_1=t_2$, this is because at this instant the oscillator states
associated with each of the qubit states are the same so that the
effect of the pulses is simply to rotate the qubit through a total
of $2\pi$. To examine the apparent coherence of the qubit we can
define the {\it envelope} of the oscillations in $P_{|+\rangle}$,
\begin{eqnarray}
E[P_{|+\rangle}(t_1+t_2)]&=&\frac{1}{2}+\frac{1}{2}{\rm e}^{-{\rm
Re}[|\alpha_0|^2\left(1-{\rm
e}^{2i\omega_1(t_1-t_2)}\right)]}\\
&=&\frac{1}{2}\left(1+{\rm
e}^{|\alpha_0|^2\left\{1-\cos(\omega_1[t_1-t_2])\right\}}\right).
\end{eqnarray}
The envelope of the oscillations is unity when $t_2=t_1$ signifying
the recoherence of the qubit. Thus we can use the echo approach to
induce recoherences in the qubit dynamics whenever we choose by
tuning $t_1(=t_2)$. We note that this particular approach also has
the advantage that inverting the state of the TLS at $t=t_1$ can
lead to an increase in the effective coherence of the TLS as
measured at $t_2\simeq t_1$ as it eliminates dephasing effects
arising from fluctuations in the TLS energy level spacings which
occur between different experimental runs\,\cite{Wallraff2007}.

\subsection{State separation and entanglement}

After evolving for a time $t$ (and without any inversion of the TLS
states), the two coherent states of the resonator that (together with
the TLS states) form a superposition are $|\alpha_{\pm}(t)\rangle$
[equation (\ref{alphapm})] and they have a separation in phase space
given by
\begin{eqnarray}
S(t)&=&\left[ (\langle X_1\rangle_{\alpha_+}-\langle
X_1\rangle_{\alpha_-})^2+(\langle X_2\rangle_{\alpha_+}-\langle
X_2\rangle_{\alpha_-})^2\right]^{1/2}\\
&=&2|\alpha_0|\sin\omega_1 t \label{sep}
\end{eqnarray}
where $\langle X_1\rangle_{\alpha_+}=\langle
\alpha_+|X_1|\alpha_+\rangle$ etc, and the phase space operators are
defined by $X_1=(a+a^{\dagger})/2$ and $X_2=(a-a^{\dagger})/(2i)$.
Because we are dealing with a pure state of the TLS and resonator we
can also obtain the entanglement of the system, $E(t)$, by
calculating the von Neumann entropy of the reduced density matrix
(of either the resonator or the TLS)\footnote{Note that the
entanglement dynamics of this system was studied very recently for
mixed states\,\cite{Clerk2}}. Evaluating this we find that it is
entirely determined by the phase space separation of the resonator
states\,\cite{mpb},
\begin{equation}
E(t)=1-\log_2\left[(1+\chi)^{(1+\chi)/2}(1-\chi)^{(1-\chi)/2}\right]
\end{equation}
where $\chi(t)=\exp(-S(t)^2/2)$. The entanglement $E(t)$ rapidly saturates at
its maximum value of unity as the separation $S$ is increased:  it reaches about
0.75 for $S=1$ and is already very close to unity for $S=2$. Note
that the decay of the qubit oscillations also depends on $S$ alone:
we can rewrite equation (\ref{p0}) as,
\begin{equation}
P_{|+\rangle}(t)=\frac{1}{2}\left(1+\chi(t)\cos\phi(t)\right)
\end{equation}
where the (real) phase is defined as
$\phi(t)=(2\Delta/\hbar+\omega_1)t+|\alpha_0|^2\sin(2\omega_1 t)$.

 The aim of these simple calculations is just to show
that the separation of the resonator states in phase space provides
an important figure of merit for the kind of experiment we have in
mind. The diameter of the `uncertainty circle'\,\cite{gerry} for a
coherent state is 1/2 and so one basic (though somewhat arbitrary)
criterion for producing a distinguishable superposition of
 resonator states is to require $S(t)\gtrsim 1$.
Although according to equation (\ref{sep}) the largest separation is
achieved when $\omega_1t=\pi/2$, in practice the limited coherence
times available for the TLS mean that the evolution times will be
such that $\omega_1 t\ll 1$ and hence we can approximate $S(t)\simeq
2|\alpha_0|\omega_1 t$. If we use the spin echo approach then the
maximum separation will be achieved at $t=t_1$ just before the TLS
state is inverted and hence to achieve a meaningful superposition we
would need to have $2|\alpha_0|\omega_1 t_1\gtrsim 1$. The details
of how a driven resonator state could be prepared in practice for
the qubit-mechanical resonator system in which the mechanical
component is formed by suspending part of the qubit circuit  is
considered in reference\ \cite{Blencowe};  we will make use of the
results obtained there when considering what kind of initial
mechanical state could be prepared in practice, but for now we point
out the crucial role played by the magnitude of the initial coherent
state, $|\alpha_0|$. The size of resonator state superposition
produced depends through $\alpha_0$ on the initial state of the
oscillator. This provides us with a way of overcoming the weak
TLS-resonator coupling and the wide separation of their time scales:
by preparing the resonator in a state with large enough $|\alpha_0|$
we can overcome the very weak interaction with the qubit to
nevertheless produce relatively large superpositions over the
relatively short times during which the TLS remains coherent. On the
other hand, unless we start with a state with non-zero $\alpha_0$
then $S$ will be zero throughout (and no entanglement will be
produced).

\section{Role of the resonator's environment}
\label{Sec:env}

We now consider the effect which the resonator's environment has on
the recoherences in the qubit. The interaction between the resonator
and its surroundings is typically modelled by including a bath of
oscillators that are weakly coupled to the resonator. This approach
is the one followed in quantum optics and although it is not clear
to what extent it represents an actual nanomechanical resonator's
environment\,\cite{schloss}, it can at least be
justified in the idealised case where dissipation in the collective
mechanical mode which forms the resonator is due only to coupling to
the bulk phonon modes in the supports of the resonator\,\cite{wr}.
In this simplified description the effects of the bath on the
`system' resonator can parameterized by a damping rate $\gamma$ for
the resonator and a temperature $T_r$, which can be expressed in
terms of the average number of the quanta the resonator would have
if it were in equilibrium with the bath,
\begin{equation}
\overline{n}=\frac{1}{{\mathrm e}^{\hbar\omega/kT_r}-1}
\label{nbar}.
\end{equation}
For sufficiently high temperatures ($k_{\rm B}T_r\gg\hbar\gamma$)
the master equation for the mechanical resonator and qubit including
the dissipative effects of the resonator's environment can be
modelled using the quantum optical damping kernel\,\cite{tannor},
\begin{equation}
\dot{\rho}=\mathcal{L}\rho=-\frac{i}{\hbar}\left[H_d,\rho\right]+\mathcal{L}_d\rho
\label{meq}
\end{equation}
where
\begin{equation}
\mathcal{L}_d\rho=-i\frac{\gamma}{2\hbar}[x,\left\{p,\rho\right\}]
-\frac{m\omega\gamma(\overline{n}+1/2)}{\hbar}[x,[x,\rho]].
\end{equation}
Assuming that the oscillator damping is very weak ($\gamma\ll
\omega$) we can further simply the dissipative part of the master
equation by using the rotating wave approximation,
\begin{equation}
\mathcal{L}_{d}\rho= -\frac{\gamma}{2}\left(a^{\dagger}a\rho+\rho
a^{\dagger}a-2a\rho a^{\dagger}\right)
-\gamma\overline{n}\left(a^{\dagger}a\rho+\rho aa^{\dagger}-a\rho
a^{\dagger}-a^{\dagger}\rho a\right). \label{rwamaster}
\end{equation}
We stress again that we use this damping kernel here to provide a
simple illustrative estimate of the dissipative dynamics of the
mechanical resonator. The true form of the mechanical damping kernel
remains somewhat uncertain and one of the aims of the experiments we
propose would be to obtain empirical information about it.

The superconducting qubit is also subject to decoherence due to
interactions with other degrees of freedom in the system apart from
the mechanical resonator\,\cite{qubitcoherence}. The dissipative
dynamics of such systems can be characterised by the relaxation
times in the equations of motion for the diagonal and
off-diagonal components of the TLS density operator. The decay of
the excited state population of the TLS is described by $T_1$, while
the decay of the TLS coherence is described by $T_2$.  In practice,
$T_1$ times have been typically an order of magnitude larger than
$T_2$ times\,\cite{Yale2005}. Since we will only consider
total evolution times $t$ (before measurement) of the system that are shorter than
$T_2$, we therefore will  always have $t\ll T_1$ and hence can neglect
relaxation of the TLS in what follows. The master equation for the
system [equation (\ref{meq})] can be written in terms of the
components $\rho_{+-}=\langle +|\rho|-\rangle$ etc, incorporating a
finite $T_2$ time as follows,
\begin{eqnarray}
\dot{\rho}_{++}&=&\mathcal{L}_{++}\rho_{++}=-i\omega_{+}[a^{\dagger}a,\rho_{++}]+{\mathcal
L}_d\rho_{++} \label{comp1}\\
\dot{\rho}_{--}&=&\mathcal{L}_{--}\rho_{--}=-i\omega_{-}[a^{\dagger}a,\rho_{--}]+{\mathcal
L}_d\rho_{--}\\
\dot{\rho}_{+-}&=&\mathcal{L}_{+-}\rho_{+-}=-\left(i2\Delta/\hbar+i\omega_1+T_2^{-1}\right)\rho_{+-} \label{comp2}\\
&&-i\omega[a^{\dagger}a,\rho_{+-}]-i\omega_1\{a^{\dagger}a,\rho_{+-}\}+{\mathcal
L}_d\rho_{+-},\nonumber
\end{eqnarray}
where $\omega_{\pm}=\omega\pm\omega_1$.

  We assume that
immediately after the first control pulse is applied to the TLS the
state of the system is given by
\begin{equation}
\rho(0)=|\psi(0)\rangle\langle\psi(0)|\otimes\rho^{(\alpha_0)}_{th},
\end{equation}
where $|\psi (0)\rangle=(|-\rangle-i|+\rangle)/\sqrt{2}$  and
$\rho^{(\alpha_0)}_{th}$ is a displaced thermal density
operator\,\cite{sato} for the resonator defined by
\begin{eqnarray}
\rho^{(\alpha_0)}_{th}&=&D(\alpha_0)\rho_{th}D^{\dagger}(\alpha_0)\\
&=&\int\int d^2\nu\frac{{\mathrm
e}^{-|\nu-\alpha_0|^2/{\overline{m}}}}{\pi\overline{m}}|\nu\rangle\langle
\nu|, \label{disp}
\end{eqnarray}
where $D(\alpha)={\rm exp}(\alpha a^{\dagger}-\alpha^*a)$ is the
displacement operator and we have defined $\nu=\alpha+\alpha_0$ in the
last line. The undisplaced thermal density operator is
\begin{equation}
\rho_{th}=\int\int d^2\alpha\frac{{\mathrm
e}^{-|\alpha|^2/{\overline{m}}}}{\pi\overline{m}}|\alpha\rangle\langle
\alpha|,
\end{equation}
where $\overline{m}=\left({{\mathrm
e}^{\hbar\omega/kT_i}-1}\right)^{-1}$. We have chosen to specify a
temperature $T_i$ for the initial state of the mechanics resonator
which can be different from that of the environment $T_r$. Simply
driving the resonator (assuming a noiseless drive) would ideally
lead to a displaced thermal state with $T_i=T_r$. However, it is
interesting conceptually to consider the case where the mechanical
resonator is somehow pre-cooled to a lower temperature than its
surroundings $T_i<T_r$. Alternatively a choice of $T_i>T_r$ provides
a simple model for the case where there is no cooling and instead
the drive adds noise to the resonator state. Although the initial
resonator state will be prepared by driving, we assume that the
drive is switched off before the first pulse is applied to the TLS.

The evolution of the component equations (\ref{comp1})-(\ref{comp2})
can be calculated very conveniently using a phase space
approach\,\cite{wandm,wands,gambetta,Clerk,serban}. The method
involves working with the Wigner transform of the components defined
as
\begin{equation}
W_{+-}(x,p;t)=\frac{1}{\hbar \pi}\int_{-\infty}^{+\infty}dy\langle
x+y|{\rho}_{+-}(t)|x-y\rangle{\rm e}^{-2ipy/\hbar}
\end{equation}
etc, which evolve according to the set of (uncoupled) partial
differential equations obtained by transforming equations
(\ref{comp1})-(\ref{comp2}). For our choice of an initial displaced
thermal state, each of the initial Wigner functions is Gaussian and
remains so during the evolution. This means that the relevant
partial differential equations for the Wigner function components
can be solved via a Gaussian ansatz. Details of the calculation
(which follows the approach used in reference \,\cite{Clerk}) are given in the
Appendix.

Using the phase space approach, we readily obtain the following
expression for $P_{|+\rangle}(t)$,
\begin{equation}
P_{|+\rangle}(t)=\frac{1}{2}+\frac{1}{2}{\rm
e}^{-t/T_2}{\rm{Re}}\left[{\rm e}^{-i2\Delta
t/\hbar+i\theta(t)}\right], \label{dp1}
\end{equation}
where
\begin{eqnarray}
\theta(t)&=&-\left(i\gamma/2+\omega_1\beta\right)t-i\ln\left[\frac{1-M}{1-M{\rm
e}^{-2i\omega_1\beta
t}}\right]\\&&-i\frac{|\alpha_0^2|}{\beta}\left({\rm
e}^{-i2\omega_1\beta t}-1\right)\left[\frac{1-M}{1-M{\rm
e}^{-2i\omega_1\beta t}}\right]\nonumber
\end{eqnarray}
with
\begin{eqnarray}
\beta&=& \left(\left[1-i\gamma/2\omega_1\right]^2-2i\gamma\overline{n}/\omega_1\right)^{1/2}\\
M&=&\frac{(2\overline{m}+1)-\beta-i\gamma/2\omega_1}{(2\overline{m}+1)+\beta-i\gamma/2\omega_1}.
\end{eqnarray}
Note that in the limit $\gamma\rightarrow 0$ we recover the much
simpler expression\,\cite{Buks}
\begin{equation}
P^{(th)}_{|+\rangle}(t)=\frac{1}{2}+\frac{1}{2}{\rm e}^{-t/T_2}{\rm
Re}\left[\frac{{\rm
e}^{-\eta(t)|\alpha_0|^2/(1+\overline{m}\eta(t))}{\rm
e}^{-i(2\Delta/\hbar+\omega_1)t}}{1+\overline{m}\eta(t)}\right],
\label{thermal2}
\end{equation}
with $\eta(t)=1-{\rm e}^{-2i\omega_1t}$.

\subsection{Echo sequence}

We now consider the case where an additional $\pi$ pulse is applied
to the system at time $t=t_1$ after the first $\pi/2$ pulse, and
then the final $\pi/2$ pulse is applied at time $t=t_f=t_1+t_2$. The
evolution of the density matrix between the two $\pi/2$ pulses can
be written as\,\cite{morigi,spinex}
\begin{equation}
\rho(t_f)={\rm e}^{{\mathcal L}(t_f-t_1)}{\mathcal R}{\rm
e}^{{\mathcal L}t_1}\rho(0),
\end{equation}
where
\begin{equation}
{\mathcal R}\rho={\rm e}^{-i\pi\overline{\sigma}_x/2}\rho{\rm
e}^{i\pi\overline{\sigma}_x/2}.
\end{equation}
 In order to calculate $P_{|+\rangle}(t_f)$ we
need the off-diagonal component of the density matrix given by,
\begin{equation}
\rho_{+-}(t_f)={\rm e}^{{\mathcal L_{+-}}t_2}\rho_{-+}(t_1)={\rm
e}^{{\mathcal L_{+-}}t_2}\rho^{\dagger}_{+-}(t_1).
\end{equation}
This evolution can again be calculated using a phase space approach
(see the Appendix for details). The resulting final probability for
finding the TLS in state $|+\rangle$ takes a very similar form to
before,
\begin{equation}
P_{|+\rangle}(t_f)=\frac{1}{2}-\frac{1}{2}{\rm
e}^{-t_f/T_2}{\rm{Re}}\left[{\rm e}^{-i2\Delta
(t_2-t_1)/\hbar+i\theta(t_f)}\right], \label{dp2}
\end{equation}
although the expression for $\theta(t_f)$ is rather complicated [it is
given in full in equation (\ref{ang2})]. It is important to note that
even though the system is damped, the phase space separation between
the components of the mechanical resonator's density matrix
corresponding to the diagonal elements of the TLS still vanishes at
$t=t_1+t_2$ for $t_2=t_1$.

The use of an echo technique allows us to filter out many of the
effects that arise just because we start with a mixed state such as
a decay in the oscillation amplitude due to averaging over the
different phases of oscillation associated with the different
resonator states in the initial mixture. The recoherence `signal'
measured at the echo time $t=2t_1$ is the {\it irreversibility} of
the system's dynamics\,\cite{morigi}. What we are in effect measuring
is the dynamics due to the resonator's damping kernel. There is no
simple way of partitioning the dissipation into a contribution from
the decoherence of spatially separated states and simple
fluctuations in the resonator's energy during the experiment: both
contribute to what is measured. An important consequence of this is
that a perfect recoherence is not achieved for $\gamma\neq 0$ even
if $\alpha_0=0$. However, when relatively large phase space
separations of the resonator state are achieved ($S\ge 1$) and the
experiment is performed on a time-scale which is very short compared
to the energy relaxation time $1/\gamma$, we can expect the
decoherence of the superposition of mechanical states to be the
dominant contribution to the irreversibility of the dynamics.

\section{Practical considerations}
\label{sec:prac}

We now turn to the question of what kinds of parameters might be
achievable in practice and hence the prospects for using the
approach we have been discussing to probe the quantum coherence of a
nanomechanical resonator in the near future. A key quantity which we
need to examine is the maximum phase space separation, $S(t_1)$,
between resonator states that can be achieved at the mid-point of an
echo experiment. As we have seen, a large initial amplitude for the
resonator $|\alpha_0|$ will enhance the phase space separation.
However, for our theoretical approach to be valid we need to ensure
not just that the parameters are achievable in practice, but also
that the approximations we made in deriving the dispersive
Hamiltonian [equation (\ref{dispersive})] remain valid.

The basic assumptions underlying our description are that the energy
scales of the TLS and the resonator and the mechanical system are
widely separated, $\hbar\omega/\Delta\ll 1$ and that we can only
expect to achieve rather weak electro-mechanical coupling,
$\kappa=\lambda/\hbar\omega\ll 1$. Furthermore, we assume that the
coherence time of the TLS in the absence of the resonator, $T_2$, is
of order $0.5~\mu$s, in line with recent experimental
results\,\cite{Yale2005,Wallraff2007} for a Cooper-pair box embedded
in a superconducting cavity. In line with this value, we assume a
maximum value of $t_1$ for the echo experiment of $\tau_c\simeq
0.2~\mu$s. For concreteness we assume a TLS energy separation
$2\Delta/h=5$~GHz and a mechanical frequency $\omega/2\pi=50$~MHz.

Within the regime where $\omega\ll \Delta/\hbar$ the maximum
amplitude of the mechanical motion for which the dispersive
Hamiltonian remains valid is set by the condition $(\lambda x/x_{zp}
\Delta)^2\ll 1$, which we can express as
$\delta=(2\kappa|\alpha_0|\hbar\omega/\Delta)^2\ll1$. We note in
passing that if $|\alpha_0|$ is small enough to satisfy this
condition then in practice it will also be small enough to ensure
that non-linear effects are unimportant in the dynamics of the
mechanical resonator\,\cite{nonlin1}.

The value of the electro-mechanical coupling constant, $\kappa$,
which can be achieved of course depends on the actual system used in
an experiment. For the specific system we have considered here
consisting of a mechanical resonator formed by suspending part of
the qubit circuit\,\cite{Blencowe}, the beam is assumed to have a width
and thickness $\simeq 200$~nm and will need to have a length of a few
microns in order to have a frequency of $50$~MHz. For such a beam
$x_{zp}\sim 10^{-14}$~m and hence we estimate\,\cite{Blencowe} that
coupling strengths up to $\kappa\simeq 0.2$ should be achievable.

The phase space separation which is achieved after a time $\tau_c$
is $S(\tau_c)\simeq2|\alpha_0|\omega_1\tau_c$ (neglecting damping of
the mechanical resonator). Using the constraint on the magnitude of
 $|\alpha_0|$, we obtain $S_{\rm
{max}}\simeq 2\pi\delta^{1/2}\kappa(\nu_m \tau_c)$. Assuming
(somewhat arbitrarily) a value of
 $\delta=0.04$,  we find that the maximum value of $\alpha_0$ that can be achieved
 without violating our assumptions will be $\simeq 5/\kappa$. Thus for
 $\tau_c=0.2~\mu$s
 and $\kappa=0.2$, we find that the maximum value of $\alpha_0$ is 25 and $S_{\rm{max}}=2.5$.
 This value for the phase space separation is encouragingly large, as the
minimum uncertainty in phase space of an oscillator state with
$\overline{n}=10$ (which corresponds to a temperature of about 25~mK
for a mechanical frequency of 50~MHz) is 2.3.

\section{Results}
\label{Sec:res}

\begin{figure}[t]
\centering \epsfig{file=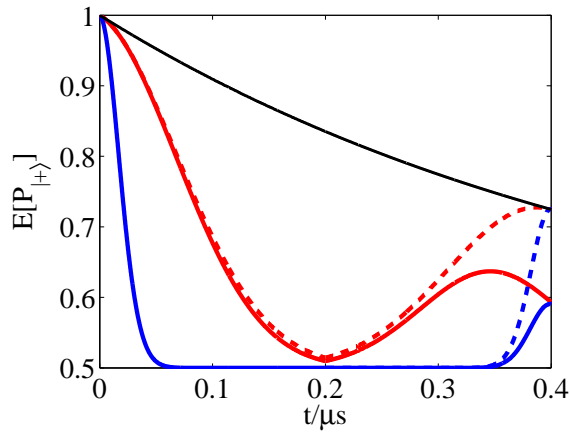,width=8cm} \caption{Envelope of
oscillations in $P_{|+\rangle}$ in an echo experiment with a $\pi$
pulse applied at $t(=t_1)=0.2~\mu$s. The blue curves are for
$\kappa=0.2$, $\overline{n}=\overline{m}=10$ and  $\alpha_0=25$. The
red curves are for the same parameters but with $\overline{m}=0$. In
each case the full curve is for $Q=3000$ and the dashed curve is for
the case without any mechanical dissipation. The black curve is the
result that would be obtained without any coupling to the mechanical
resonator.} \label{fig:f1}
\end{figure}

We now use the results of the previous section to explore the
behaviour of the oscillations in $P_{|+\rangle}$ during an echo
experiment using practicable values of all the parameters. We start
by examining the envelope of the oscillations in $P_{|+\rangle}$
during an echo experiment before and after an inversion pulse at
$t=t_1$. The envelope of the oscillations is defined by
\begin{equation}
E[P_{|+\rangle}(t)]=\frac{1}{2}+\frac{1}{2}{\rm e}^{-t/T_2}{\rm
e}^{-{\rm{Im}}[\theta(t)]}. \label{dpe}
\end{equation}
where $\theta(t)$ is given by equation (\ref{dp1}) for times $t<t_1$
and by equation (\ref{ang2}) for $t>t_1$. An example of the expected
behaviour as a function of $t$ is shown in figure \ref{fig:f1}. We
assume throughout the parameter values discussed in the previous
section ($\omega/2\pi=50$~MHz, $\nu_a=5$~GHz and $T_2=0.5~\mu$s) and
consider the maximal coupling $\kappa=0.2$ and amplitude
$\alpha_0=25$. The strength of the mechanical dissipation is
specified by the resonator's Q-factor, $Q=\omega/\gamma$. We have
taken $\overline{n}=10$ and as well as considering the case where
$\overline{m}=\overline{n}$, we also (for theoretical interest)
consider the extreme case where the resonator is somehow pre-cooled
to its ground state, $\overline{m}=0$.

From the curves in figure \ref{fig:f1} we can see that the
mechanical resonator is likely to have a strong effect on the TLS.
It is interesting to compare the curves with and without the
inclusion of a finite Q-factor for the mechanical resonator. In an
echo experiment, only mechanical dissipation leads to a deviation
from the uncoupled value of $E[P_{|+\rangle}]$ at the echo point,
$t=2t_1$ (i.e.,\ the recoherence). Although an initial mixture of
resonator states leads to an average over phases associated with
each of the different states and hence a strong enhancement of the
apparent dephasing of the TLS during the first part of the
experiment ($t\leq t_1$), after the echo each of these phases
unwinds and hence they do not affect the behaviour at $t=2t_1$. On the other hand, when
dissipation is included we see that the echo signal can be
substantially reduced.

It is important to note that dissipation of the mechanical resonator
has only a very small effect on the behaviour of the signal
$E[P_{|+\rangle}]$ before the $\pi$ pulse is applied. This is
because the decay of this signal is dominated by the separation  of
the resonator states and the averaging over the different phases
associated with each of the states in the thermal mixture. The
decoherence of the mechanical resonator only starts to occur once a
superposition has been produced and by the time it has started to
develop, the value of $E[P_{|+\rangle}]$ is already close to 0.5.
Thus, the decoherence of the mechanical resonator can only really be
measured by using the echo signal around $t\simeq 2t_1$.

\begin{figure}[t]
\centering \epsfig{file=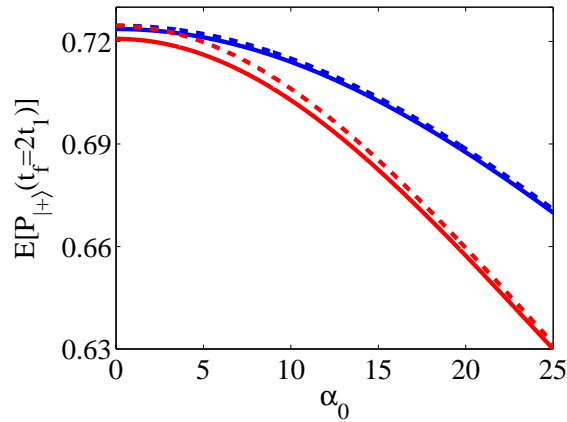,width=8cm} \caption{Envelope of
oscillations in $P_{|+\rangle}$ in an echo experiment measured at
time $t_f=2t_1$ as a function of $\alpha_0$. The full (dashed)
curves are for $\overline{m}=\overline{n}$ ($\overline{m}=0$) with
$\kappa=0.2$, $t_1=0.2~\mu$s and $Q=10^4$. The red curves are for
$\overline{n}=20$ and the blue curves are for $\overline{n}=10$.}
\label{fig:f4}
\end{figure}

It is interesting to note that pre-cooling the resonator does not
affect the echo signal by very much. This is again because the phase
averaging that occurs for a mixed state is largely removed by the
use of the echo sequence. However, in the presence of dissipation
the states involved in a thermal mixture will have slightly
different amplitudes (compared to the average $\alpha_0$) and hence
will all be affected slightly differently by the coupling to the
environment during the evolution: the mixed initial state
curve ($\overline{m}=\overline{n}$) does not exactly match the
pre-cooled (pure) one ($\overline{m}=0$) at $t=2t_1$. This behaviour
can be seen more clearly in figure \ref{fig:f4} which focusses on
the echo signal at $t=2t_1$ for a range of $\alpha_0$ values. Over
the relatively short time of the echo $t_1\ll 1/\gamma=Q/\omega$,
energy diffusion is a very weak effect and hence the evolution of
the thermal state is very similar to an average over pure initial
states with a range of $\alpha_0$ values ($\sim
\overline{m}^{1/2}$). Thus the results for the initially mixed
($\overline{m}=\overline{n}$) and pure states ($\overline{m}=0$)
become very close for larger $\alpha_0$ values where the variation
of the envelope signal with $\alpha_0$ is approximately linear (on a
scale $\sim \overline{m}^{1/2}$), and overall the curves are closer
for lower $\overline{n}$.

\begin{figure}[t]
\centering \epsfig{file=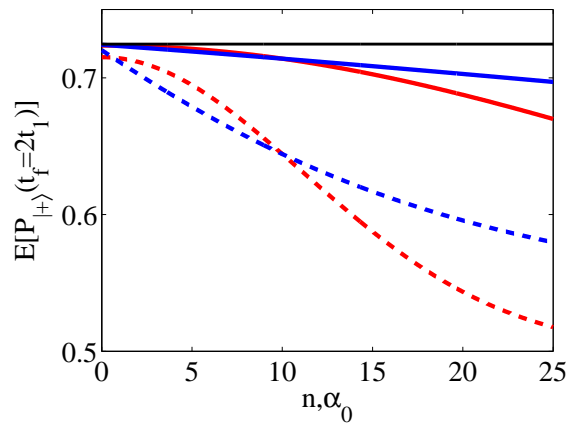,width=8cm} \caption{Envelope
of oscillations in $P_{|+\rangle}$ in an echo experiment with a
$\pi$ pulse applied at $t(=t_1)=0.2~\mu$s, measured at time
$t_f=2t_1$. The blue curves are for $\kappa=0.2$, with $\alpha_0=10$
and $\overline{n}=\overline{m}$ varied from 0 to 25. The red curves
are for the same parameters but with $\overline{n}=\overline{m}=10$
and $\alpha_0$ varied from 0 to 25. In each case the dashed curve is
for $Q=10^3$ and the full curve is for $Q=10^4$. The black line is
the result that would be obtained without any dissipation to the
mechanical resonator.} \label{fig:f2}
\end{figure}

In figure \ref{fig:f2}, we compare the effects of varying the
temperature of the mechanical resonator's environment and the
amplitude of the initial state on the echo signal at $t=2t_1$.
Increasing the value of either $\overline{n}$ or $\alpha_0$ reduces
the recoherence at the echo, but the dependences are rather
different. An important part of any experiment would be to test this
behaviour, something which could readily be done for $\alpha_0$ by
simply varying the initial drive applied to the mechanical resonator
to prepare it in states of different amplitude.

\begin{figure}[t]
\centering \epsfig{file=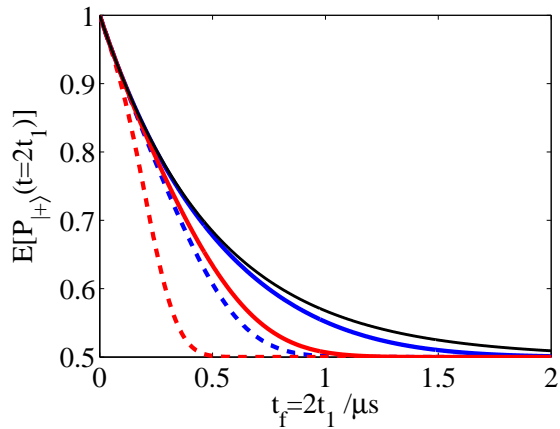,width=8cm} \caption{Envelope of
oscillations in $P_{|+\rangle}$ in an echo experiment with a $\pi$
pulse applied at $t=t_1$, measured at time $t_f=2t_1$ as a function
of $t_f$. The full (dashed) curves are for $\kappa=0.1$
($\kappa=0.2$), with $\alpha_0=25$ and
$\overline{n}=\overline{m}=10$. The red curves are for $Q=10^3$ and
the blue curves are for $Q=10^4$. The black line is the result that
would be obtained without any dissipation to the mechanical
resonator.} \label{fig:f3}
\end{figure}

Finally, in figure \ref{fig:f3} we explore how changing the time
between the pulses $t_1$ (and hence the total time for the echo
experiment $t_f=2t_1$) affects the behaviour at the echo point. This
plot shows clearly the strong deviation from simple exponential
decay that the coupling to the resonator can lead to. As we have
already discussed, the superposition of resonator states takes time
to develop and hence it takes a while before decoherence of the
mechanical resonator can start to affect the dynamics of the TLS
which is measured;   all the curves in figure \ref{fig:f3}
initially lie very close to each other. However, at longer times the
dissipative effect of the mechanical system's environment starts to
have an important influence. Furthermore, it is clear that for
strong enough coupling the decay of $E[P_{|+\rangle}(t=2t_1)]$
occurs on a much faster scale than the relaxation of the resonator's
energy, $\gamma=\omega/Q$, a clear sign that it is the loss of the
mechanical system's quantum coherence which drives the process.

The range of $Q$ factors which we have used here, $10^3-10^4$, is
appropriate for a resonator formed by a suspended metal
film\,\cite{metalfilm}. However, where the resonator consists of a
semiconductor beam which is then coated in a metal layer, somewhat
higher Q-factors can occur~\cite{Naik} (up to $\sim 10^5$). For very
high Q-factor resonators the amplitude of the echo signal will be
completely dominated by the qubit decoherence and the contribution
from the resonator's bath may eventually become too small to measure
in practice. In this regime the measurement of the qubit
recoherences would only allow an upper bound for the decoherence of
the mechanical system to be established.

\section{Conclusions and Discussion}
\label{Sec:conc}

In this paper we have discussed how a superconducting qubit can be
used to probe the quantum coherence of a nanomechanical resonator
using methods very similar to those applied in recent optical cQED
experiments. In particular, we explored how an echo experiment could
be used to systematically explore the quantum dynamics of a
mechanical resonator using a superconducting qubit tuned to the
degeneracy point as a probe.

The advantages of the echo approach go beyond the practicalities of
the system. The ability to control the duration of the experiment
and to vary the separation of resonator states produced (by varying
the initial amplitude $\alpha_0$) will make it much easier to draw
strong conclusions about the nature of the mechanical system's
environment. Interestingly, we found that over a range of
temperatures (corresponding to thermal occupation numbers of the
resonator up to  $\sim 20$) the recoherences were likely to be
affected only very weakly by the variance of the initial resonator
state implying that it is by no means necessary to prepare the
resonator in a pure state to obtain important information about its
quantum dynamics. We expect the echo technique to be rather robust
in the sense that it should give useful information about the
quantum coherence of the resonator for a rather wide range of
parameters. The larger the separation of states achieved during an
echo experiment, the more the magnitude of the recoherences will
tell us about the coherence properties of the mechanical system.
However, there is no threshold below which nothing is learnt: even
if only a very small separation ($S\leq 1$) is achieved then some
information is nevertheless obtained about the dissipative dynamics
of the mechanical resonator beyond just the energy relaxation rate.

Since a great deal will be inferred from the deviations between the measured
dynamics and the reversible dynamics calculated using the dispersive
Hamiltonian, it will in practice be necessary to be able to
discriminate between contributions arising from the resonator's
environment and those due to the inevitable corrections to the model
Hamiltonian which is an approximate form. Therefore, an important
future extension of the current work would be to carry out a
systematic numerical study of the coupled qubit-resonator dynamics
using the full Hamiltonian of the system. Such an approach would not
just allow us to calculate the effects of corrections to the
dispersive Hamiltonian, but also allow a more comprehensive modelling of the
qubit's environment to include energy relaxation. As recent
experiments\,\cite{Wallraff2007} have begun to approach the regime
where $T_2>T_1$, the inclusion of a finite $T_1$ is becoming
increasingly relevant.

\section*{Acknowledgments}

We thank E. Buks and G. Milburn for useful discussions. This work
was supported financially by the EPSRC under grant EP/E03442X/1
(ADA), by the NSF under NIRT grant CMS-0404031 and by the
Foundational Questions Institute (MPB).

\appendix

\section{Calculation of TLS decoherence for a damped resonator}

In this Appendix we calculate the dynamics of the Wigner function
component $W_{+-}$ including the effects of the environment. We
start from the equation of motion for $\rho_{+-}$ [equation
(\ref{comp2})], which in terms of the interaction picture,
\begin{equation}
\tilde{\rho}_{+-}(t)={\rm e}^{[i2\Delta /\hbar+1/T_2]t}\rho_{+-}(t),
\end{equation}
becomes
\begin{equation}
\dot{\tilde{\rho}}_{+-}=-{i}\left[\omega
a^{\dagger}a,\tilde{\rho}_{+-}\right] -i\omega_1\left\{
a^{\dagger}a+1/2,\tilde{\rho}_{+-}\right\}+{\mathcal
L}_d\tilde{\rho}_{+-}. \label{evol}
\end{equation}
Defining the Wigner transform in the usual way,
\begin{equation}
\tilde{W}_{+-}(x,p;t)=\frac{1}{\hbar
\pi}\int_{-\infty}^{+\infty}dy\langle
x+y|\tilde{\rho}_{+-}(t)|x-y\rangle{\rm e}^{-2ipy/\hbar}
\end{equation}
we obtain the Wigner-transformed equation of motion,
\begin{eqnarray}
\frac{\partial \tilde{W}_{+-}}{\partial
t}&=&\left[\frac{\gamma}{2}\tilde{x}-\omega\tilde{p}\right]\frac{\partial
\tilde{W}_{+-}}{\partial
\tilde{x}}+\left[\frac{\gamma}{2}\tilde{p}+\omega\tilde{x}\right]\frac{\partial
\tilde{W}_{+-}}{\partial
\tilde{p}}+\gamma(\overline{n}+1/2)\left(\frac{\partial^2
\tilde{W}_{+-}}{\partial \tilde{x}^2}+\frac{\partial^2
\tilde{W}_{+-}}{\partial \tilde{p}^2}\right)\nonumber\\
&& +\gamma
\tilde{W}_{+-}-i\frac{\omega_1}{2}\left[\tilde{x}^2+{\tilde
p}^2-\frac{\partial^2 \tilde{W}_{+-}}{\partial
\tilde{x}^2}-\frac{\partial^2 \tilde{W}_{+-}}{\partial
\tilde{p}^2}\right],\label{wpde}
\end{eqnarray}
where $\tilde{x}=x/x_{zp}$ and $\tilde{p}=p(2/m\hbar\omega)^{1/2}$.

In order to solve this equation of motion we make a {\it Gaussian
ansatz}, assuming that the Wigner function takes the form of a
Gaussian multiplied by a phase factor
\begin{equation}
\tilde{W}_{+-}(\tilde{x},\tilde{p};t)=W_G(\tilde{x},\tilde{p};t)\frac{{\rm
e}^{i\theta'}}{2}=\frac{{\rm
e}^{\frac{-1}{2D}\left[\sigma_p(\tilde{x}-\overline{x})^2
-2\sigma_{xp}(\tilde{x}-\overline{x})(\tilde{p}-\overline{p})+\sigma_x(\tilde{p}-\overline{p})^2\right]}}{2\pi\sqrt{D}}\frac{{\rm
e}^{i\theta'}}{2} \label{wgdefn}
\end{equation}
where $D$ is the determinant of the matrix
\begin{equation}
\left(\begin{array}{cc} \sigma_x & \sigma_{xp}\\
\sigma_{xp}& \sigma_p\end{array}\right), \label{det}
\end{equation}
and the five parameters
$(\overline{x},\overline{p},\sigma_x,\sigma_p,\sigma_{xp})$ and the
phase $\theta$ are taken to be time dependent. Defined in this way
$W_G(\tilde{x},\tilde{p})$ is normalized (i.e.\ integrating it over
all $\tilde{x},\tilde{p}$ values gives unity) and so the factor of
${\rm e}^{i\theta'(t)}/2$ has been introduced as ${\rm
Tr}[\rho_{+-}(t)]$ is by definition (for a TLS) a complex number
with amplitude $\leq 1/2$. The initial Gaussian remains a Gaussian
for all times (albeit with different parameters) and hence remains
normalized, thus
\begin{equation}
{\rm Tr}\left[\tilde{\rho}_{+-}(t)\right]=\int d\tilde{p}\int
d\tilde{x} W_G(\tilde{x},\tilde{p};t)\frac{{\rm
e}^{i\theta'(t)}}{2}=\frac{{\rm e}^{i\theta'(t)}}{2}
\end{equation}
and hence
\begin{equation}
{\rm Tr}\left[{\rho}_{+-}(t)\right]={\rm e}^{-i2\Delta
t/\hbar-t/T_2}\frac{{\rm e}^{i\theta'(t)}}2.
\end{equation}
This function is all we need to calculate the probability of finding
the qubit in state $|+\rangle$,
\begin{equation}
P_{|+\rangle}(t)=\frac{1}{2}\left(1-2{\rm Im}\left\{{\rm
Tr}[\rho_{+-}(t)]\right\}\right).
\end{equation}
Thus using the definition of the initial state of the TLS [equation
\ref{psi0}] we can see that $\theta'(0)=3\pi/2$.

In principal we can solve for the time dependence of the six
parameters in the Gaussian by substituting the ansatz into the
equation of motion directly\,\cite{tannor} and equating powers of
$\tilde{x},\tilde{p}$. However, in practice the problem is more
readily solved\,\cite{Clerk,serban} using the {\it characteristic
function} which is defined by the relation\,\cite{statmech}
\begin{eqnarray}
G(k,q)&=&\int d\tilde{x}\int d\tilde{p} \tilde{W}_{+-}(\tilde{x},\tilde{p};t) {\rm e}^{ik\tilde{x}}{\rm e}^{iq\tilde{p}}\\
&=&\frac{{\rm e}^{i\theta'}}{2}{\rm
e}^{i(k\overline{x}+q\overline{p})-(k^2\sigma_x+q^2\sigma_p+2kq\sigma_{xp})/2}.
\label{trial}
\end{eqnarray}
The equation of motion for the characteristic function is readily
derived from the corresponding one for the Wigner function,
\begin{eqnarray}
\frac{\partial G}{\partial t}&=&\left(\omega k-\gamma
q/2\right)\frac{\partial G}{\partial q}-\left(\omega q+\gamma
k/2\right)\frac{\partial G}{\partial k}\\
&&-\left[\gamma(\overline{n}+1/2)+i\omega_1/2\right]\left(k^2+q^2\right)G+\frac{i\omega}{2}
\left(\frac{\partial^2 G}{\partial k^2}+\frac{\partial^2 G}{\partial
q^2}\right)\nonumber
\end{eqnarray}
Substituting the trial function into the left-hand side of this
equation and equating powers of $k,q,kq$ etc, leads directly to a
set of equations of motion for the six time dependent parameters,
\begin{eqnarray}
\dot{\theta}&=&-\frac{\omega_1}{2}\left[\overline{x}^2+\overline{p}^2+\sigma_x+\sigma_p\right]\\
 \dot{\overline{
p}}&=&-\omega\overline{x}-\gamma\overline{p}/2-i\omega_1(\sigma_{xp}\overline{x}+\sigma_p\overline{p})\\
\dot{\overline{
x}}&=&\omega\overline{p}-\gamma\overline{x}/2-i\omega_1(\sigma_{xp}\overline{p}+\sigma_x\overline{x})\\
\dot{\sigma}_{xp}&=&\omega(\sigma_p-\sigma_x)-\gamma\sigma_{xp}-i\omega_1\sigma_{xp}(\sigma_p+\sigma_x)\\
\dot{\sigma}_{x}&=&2\omega\sigma_{xp}-\gamma[\sigma_{x}-N]-i\omega_1(\sigma_x^2+\sigma_{xp}^2-1)\\
\dot{\sigma}_{p}&=&-2\omega\sigma_{xp}-\gamma[\sigma_{p}-N]-i\omega_1(\sigma_p^2+\sigma_{xp}^2-1),
\end{eqnarray}
where we have defined $N=2\overline{n}+1$. We now need to solve
these equations subject to appropriate initial conditions.

Assuming a thermal state displaced by the coherent amplitude
$\alpha_0$, the set of initial conditions is as follows:
$\overline{x}(0)=2{\rm Re}[\alpha_0]$, $\overline{p}(0)=2{\rm
Im}[\alpha_0]$, $\sigma_x(0)=\sigma_p(0)=2\overline{m}+1$ and
$\sigma_{xp}(0)=0$. With these initial conditions it is clear that
$\sigma_{xp}$ will remain zero for all times and the position and
momentum variances will always remain the same,
$\sigma_x(t)=\sigma_p(t)=1+\sigma(t)$, following the simplified
equation
\begin{equation}
\dot{\sigma}=-\gamma[\sigma-\sigma_0]-i\omega_1(2\sigma+\sigma^2)
\label{sigma}
\end{equation}
where $\sigma_0=\sigma(0)=2\overline{m}$. The solution of this
equation gives
\begin{equation}
\sigma_x(t)=\sigma_p(t)=1+\sigma(t)=i\frac{\gamma}{2\omega_1}+\beta\left[
\frac{1+M{\rm e}^{-2i\omega_1\beta t}}{1-M{\rm e}^{-2i\omega_1\beta
t}}\right]
\end{equation}
where
\[
M=\frac{(2\overline{m}+1)-\beta-i\gamma/2\omega_1}{(2\overline{m}+1)+\beta-i\gamma/2\omega_1}
\]
and
\[
\beta=\left[\left(1-\frac{i\gamma}{2\omega_1}\right)^2-\frac{2i\gamma\overline{n}}{\omega_1}\right]^{1/2}.
\]

The final part of the calculation involves calculating
$\overline{x}(t)$ and $\overline{p}(t)$ and hence obtaining the
phase $\theta(t)$. The equations for the averages are most easily
solved in terms of the variables
$a_{1(2)}=(\overline{x}+(-)i\overline{p})/2$ which obey the
equations of motion
\begin{eqnarray}
\dot{a}_1&=&\left(-i\omega
-\frac{\gamma}{2}-i\omega_1(1+\sigma(t))\right)a_1 \label{a1}\\
\dot{a}_2&=&\left(i\omega
-\frac{\gamma}{2}-i\omega_1(1+\sigma(t))\right)a_2 \label{a2}
\end{eqnarray}
and can be integrated to give,
\begin{eqnarray}
a_1(t)&=&a_1(0){\rm e}^{(-i\omega-\gamma/2)t}{\rm
e}^{-i\omega\int_0^t[1+\sigma(t')]dt'}\\
a_2(t)&=&a_2(0){\rm e}^{(i\omega-\gamma/2)t}{\rm
e}^{-i\omega\int_0^t[1+\sigma(t')]dt'}.
\end{eqnarray}
The integral in the exponentials is readily evaluated,
\begin{equation}
\int_0^t[1+\sigma(t')]dt'=\left(\frac{i\gamma}{2\omega_1}+\beta\right)t+\frac{1}{i\omega_1}\ln\left[\frac{1-M{\rm
e}^{-2i\omega_1\beta t}}{1-M}\right],
\end{equation}
and hence we find
\begin{equation}
a_{1(2)}(t)=a_{1(2)}(0){\rm
e}^{(-(+)i\omega-i\omega_1\beta)t}\left[\frac{1-M}{1-M{\rm
e}^{-2i\omega_1\beta t}}\right].
\end{equation}
The initial values of $a_{1(2)}$ are
$a_{1}(0)=\alpha_0,a_{2}(0)=\alpha^*_0$.

Finally then we are in a position to obtain the required phase,
$\theta'(t)$. Noting that $\overline{x}^2+\overline{p}^2=4a_1a_2$
and using the appropriate initial condition ($\theta'(0)=3\pi/2$),
we obtain
\begin{eqnarray}
\theta'(t)&=&\theta'(0)-\omega_1\int_0^t(1+\sigma(t'))dt'-2\omega_1\int_0^ta_1(t')a_2(t')dt'\\
&=&3\pi/2-\left(i\frac{\gamma}{2}+\omega_1\beta\right)t
-i\ln\left[\frac{1-M}{1-M{\rm e}^{-i2\omega_1\beta
t}}\right]\\&&-i\frac{|\alpha_0|^2}{\beta}(1-M)\left[\frac{{\rm
e}^{-i2\omega_1\beta t}-1}{1-M{\rm e}^{-i2\omega_1\beta
t}}\right].\nonumber
\end{eqnarray}
Thus we arrive at our final result,
\begin{equation}
{\rm Tr}[\rho_{+-}(t)]=\left(\frac{-i}{2}\right){\rm e}^{-i2\Delta
t/\hbar-t/T_2+i\theta(t)}, \label{apen}
\end{equation}
where we have defined $\theta(t)=\theta'(t)-\theta'(0)$. This result
[equation \ref{apen}] and the expression for $\theta'(t)$ above
gives equation (\ref{dp1}) in the main text.

We now extend this calculation to consider the spin-echo case where
the system is prepared and allowed to evolve in the way we have been
considering, but after time $t_1$ an additional control pulse is
applied to invert the populations of the two eigenstates. The system
is then allowed to evolve for a further time $t_2$ before a final
control pulse is applied and then a measurement is made.

In order to obtain $\rho_{+-}(t_f=t_1+t_2)$ we need to solve
equation (\ref{evol}) twice: first for the period $t_1$ and then
using the Hermitian conjugate of this solution as the initial
condition for a further evolution over time $t_2$. As before we use
the Wigner function approach and hence use $[\tilde{W}_{+-}(t_1)]^*$
as an initial condition for equation (\ref{wpde}).

Using the above calculation we can immediately write down
\begin{equation}
W^*_{+-}(\tilde{x},\tilde{p};t_1)=\frac{{\rm
e}^{i\phi'(t_1)}}{2}W_G(\tilde{x},\tilde{p};t_1),
\end{equation}
where the Gaussian Wigner function is in this case parameterized by
\begin{eqnarray}
\sigma_x(t_1)&=&\sigma_p(t_1)=\sigma_1(t_1)+1\\
\sigma_1(t_1)&=&-1-i\frac{\gamma}{2\omega_1}+\beta^*\left[\frac{1+M^*{\rm
e}^{i2\omega_1\beta^*t_1}}{1-M^*{\rm
e}^{i2\omega_1\beta^*t_1}}\right]\label{set1}\\
a_1(t_1)&=&\alpha^*_0{\rm
e}^{i(\omega+\omega_1\beta^*)t_1}\left[\frac{1-M^*}{1-M^*{\rm
e}^{i2\omega_1\beta^*t_1}}\right]\\
a_2(t_1)&=&\alpha_0{\rm
e}^{-i(\omega-\omega_1\beta^*)t_1}\left[\frac{1-M^*}{1-M^*{\rm
e}^{i2\omega_1\beta^*t_1}}\right]\label{set2}
\end{eqnarray}
with
\begin{eqnarray}
\phi'(t_1)&=&-3\pi/2-\left(\frac{i\gamma}{2}-\omega_1\beta^*\right)t_1-i\ln\left[\frac{1-M^*}{1-M^*{\rm
e}^{i2\omega_1\beta^*t_1}}\right]\\&&-i\frac{|\alpha_0|^2}{\beta^*}(1-M^*)\left[\frac{{\rm
e}^{i2\omega_1\beta^*t_1}-1}{1-M^*{\rm
e}^{i2\omega_1\beta^*t_1}}\right].\nonumber
\end{eqnarray}

The final step is then to use $W^*_{+-}(t_1)$ as the initial
condition for $W_{+-}$ evolved over a time $t_2$. Solving the
equations of motion for the Gaussian parameters [equations
(\ref{sigma}), (\ref{a1}) and (\ref{a2})] using the initial conditions
given by equations (\ref{set1})-(\ref{set2}) above we finally obtain
the phase parameter which is used in equation~(\ref{dpe}) for
$t>t_1$,
\begin{eqnarray}
\theta(t_f)&=&(\phi'(t_1)+3\pi/2)-\left(\frac{i\gamma}{2}+\omega_1\beta\right)t_2-i\ln\left[\frac{1-M'}{1-M'{\rm
e}^{-i2\omega_1\beta
t_2}}\right]\nonumber\\&&-i\frac{a_1(t_1)a_2(t_1)}{\beta
}(1-M')\left[\frac{{\rm e}^{-i2\omega_1\beta t_2}-1}{1-M'{\rm
e}^{-i2\omega_1\beta t_2}}\right] \label{ang2}
\end{eqnarray}
where
\begin{equation}
M'=\frac{\sigma_1(t_1)+\left(1-\frac{i\gamma}{2\omega_1}\right)-\beta}{\sigma_1(t_1)+\left(1-\frac{i\gamma}{2\omega_1}\right)+\beta}.
\end{equation}


\section*{References}
\begin{thebibliography}{99}
\bibitem{ABS} Armour AD, Blencowe MP and Schwab KC (2002) {\it Phys.
Rev. Lett.} {\bf 88} 148301.
\bibitem{Buks}Buks E and Blencowe MP (2006)  {\it Phys. Rev. B} {\bf 74},
174504.
\bibitem{Tian} Tian L (2005) {\it Phys. Rev. B} {\bf 72} 195411.
\bibitem{deconstruct}  Anglin JR, Paz JP and Zurek WH (1997)  {\it Phys. Rev. A}
{\bf 55} 4041.
\bibitem{ramsey} Brune M, Hagley E, Maitre X, Maali A,  Raimond JM
and Haroche S (1996) {\it Phys. Rev. Lett.} {\bf 77} 4887.
\bibitem{spinex} Meunier T, Gleyzes S, Maioli P, Auffeves A,
Nogues G, Brune M, Raimond JM and Haroche S (2005) {\it Phys. Rev.
Lett.} {\bf 94} 010401.
\bibitem{Haroche}Haroche S and Raimond JM (2006) {\it Exploring the
Quantum}, Oxford University Press Oxford UK.
\bibitem{monroe} Monroe CD, Meekhof DM,  King BE and
Wineland DJ (1996) {\it Science} {\bf 272} 1131.
\bibitem{leibfried} Leibfried D, Blatt R, Monroe C and
Wineland D (2003) {\it Rev. Mod. Phys.} {\bf 44} 281.
\bibitem{mcdonnell} McDonnell MJ, Home JP, Lucas DM, Imreh G,
Keitch BC, Szwer DJ, Thomas NR, Webster SC, Stacey DN and Steane AM
(2007) {\it Phys. Rev. Lett.} {\bf 98} 063603
\bibitem{Yale1} Wallraff A, Schuster DI, Blais A, Frunzio L, Huang
R-S, Majer J, Kumar S, Girvin SM and Schoelkopf RJ (2004) Nature
{\bf 431} 162
\bibitem{phase} Bertet P, Chiorescu I, Burkard G, Semba K,
Harmans CJPM, DiVincenzo DP and Mooj JE (2005) {\it Phys. Rev.
Lett.} {\bf 95} 257002.
\bibitem{mpb} Blencowe MP (2004) {\it Phys. Rep.} {\bf 395} 159.
\bibitem{leggett} Leggett AJ (2002) {\it J. Phys. Cond. Matt.} {\bf 14}
R415.
\bibitem{Cleland} Cleland AN and Geller MR (2004) {\it Phys. Rev.
Lett.}
{\bf 93} 070501.
=
\bibitem{Blencowe} Blencowe MP and Armour AD (2008)
(unpublished).
\bibitem{pert} Irish EK and Schwab KC (2003) {\it Phys. Rev. B}
{\bf 68} 155311.
\bibitem{ringsmuth} Ringsmuth AK and Milburn GJ (2007) {\it J. Mod.
Optics} {\bf 54} 2223.
\bibitem{qubitcoherence} Ithier G, Collin E, Joyez P, Meeson PJ,  Vion D, Esteve
D, Chiarello F, Makhlin Y, Schriefl J and Sch\"{o}n G (2005) {\it
Phys. Rev. B} {\bf 72} 134519.
\bibitem{schloss} Schlosshauer M, Hines AP and Milburn GJ (2008) {\it
Phys. Rev. A} {\bf 77} 022111.
\bibitem{flux1}  Zhou X and Mizel A (2006) {\it Phys. Rev. Lett.} {\bf 97} 267201; Xue
F, Wang Y, Sun CP, Okamoto H, Yamaguchi H and Semba K (2007) {\it
New J. Phys.} {\bf 9} 35.
\bibitem{Yale2005} Wallraff A, Schuster DI, Blais A, Frunzio L,
Majer J, Devoret MH, Girvin SM and Schoelkopf RJ (2005) {\it Phys.
Rev. Lett.} {\bf 95} 060501.
\bibitem{Wallraff2007} Leek PJ, Fink JM, Blais A, Bianchetti R,
G\"{o}ppl M, Gambetta JM, Schuster DI, Frunzio L, Schoelkopf RJ and
Wallraff A (2007) {\it Science} {\bf 318} 1889.
\bibitem{Huang}Huang XMH, Zorman CA, Mehragany M and Roukes ML (2003) {\it Nature} {\bf 421} 496.
\bibitem{Naik} Naik A, Buu O, LaHaye MD, Armour AD, Clerk AA, Blencowe MP and Schwab KC (2006) {\it Nature} {\bf
443} 193.
\bibitem{larson} Larson J and Stenholm S (2006) {\it Phys. Rev. A}
{\bf 73} 033805.
\bibitem{messiah} Messiah A (1962) {\it Quantum Mechanics}
North-Holland Amsterdam.

\bibitem{graham}Graham R and H\"{o}hnerbach M (1984) {\it Z. Phys. B} {\bf
57} 233.
\bibitem{gambetta} Gambetta J, Blais A, Schuster DI,
Wallraff A, Frunzio L, Majer J, Devoret MH, Girvin SM and Schoelkopf
RJ (2006) {\it Phys. Rev. A} {\bf 74} 042318.
\bibitem{Clerk} Clerk AA and Wahyu Utami D (2007) {\it Phys. Rev. A} {\bf
75} 042302.
\bibitem{Buksqnd} Buks E, Arbel-Segev E, Zaitsev S, Abdo B and
Blencowe MP (2008) {\it Europhys. Lett.} {\bf 81} 10001


\bibitem{Blais}Blais A,  Huang RS, Wallraff A, Girvin SM and Schoelkopf RJ (2004) {\it Phys. Rev. A} {\bf 69}
062320.
\bibitem{gerry} Gerry CC and Knight PL (2005) {\it Introductory Quantum
Optics} Cambridge Univesrity Press, Cambridge UK.
\bibitem{vion} Vion D, Aassime A, Cottet A, Joyez P, Pothier H, Urbina C, Esteve D and
Devoret M H (2002) {\it Science} {\bf 296} 886
\bibitem{Clerk2} Wahyu Utami D and Clerk AA  (2008)  arXiv:0803.0541
(unpublished).
\bibitem{wr} Wilson-Rae I (2007) arXiv:0710.0200 (unpublished).
\bibitem{tannor} Kohen D, Marston CC and Tannor DJ (1997) {\it J. Chem.
Phys.} {\bf 107} 5236.
\bibitem{sato} Saiyo H and Hyuga H (1996) {\it J. Phys. Soc. Japan} {\bf 65}
1648.
\bibitem{wandm} Walls DF and Milburn GJ (1985) {\it Phys. Rev. A}
{\bf 31} 2403.
\bibitem{wands} Savage CM and Walls DF (1985)  {\it Phys. Rev. A}
{\bf 32} 2316.
\bibitem{serban}  Serban I, Solano E and  Wilhelm FK (2007) {\it Europhys. Lett.} {\bf 80} 40011.
\bibitem{morigi}  Morigi G, Solano E, Englert B-G and Walther H
(2002) {\it Phys. Rev. A} {\bf 65} 040102
\bibitem{nonlin1} Postma HWC, Kozinsky I, Husain A and Roukes ML (2005) {\it Appl. Phys. Lett.} {\bf 86}
223105
\bibitem{metalfilm}Li TF, Pashkin Yu P, Astafiev O, Nakamura Y, Tsai
JS and Im H (2008) {\it Appl. Phys. Lett.} {\bf 92} 043112
\bibitem{statmech} See for example, Appendix 2 of Zwanzig R (2001) {\it
Nonequilibrium Statistical Mechanics} Oxford University Press Oxford
UK.
\end{thebibliography}
\end{document}